\documentclass[10pt,conference]{IEEEtran}
\usepackage{amsmath}
\usepackage{xcolor}               
\usepackage[switch]{lineno}

\usepackage{cite}
\usepackage[table]{xcolor}
\usepackage{hyperref}
\usepackage{float}
\usepackage{multirow}
\usepackage{makecell}
\usepackage{arydshln}
\usepackage{enumitem}
\usepackage{xspace}
\usepackage{textcomp}
\usepackage{url}
\usepackage{fix-cm}
\usepackage[breakable]{tcolorbox}
\usepackage{balance}
\usepackage{tikz}

\newcommand{\solidcircle}[1]{%
  \tikz[baseline=(char.base)]{
    \node[shape=circle, draw, fill=black, inner sep=0.5pt] (char) {\textcolor{white}{\textbf{#1}}};
  }%
}
\newcommand{\figmargin}{\vspace{-12pt}}
\newcommand{\tabmargin}{\vspace{-8pt}}
\newcommand{\secmargin}{\vspace{-2pt}}
\newcommand{\subsecmargin}{\vspace{-2pt}}
\newcommand{\boxmargin}{1mm}
\newcommand{\toolname}{\textsc{DrainCode}\xspace} 

\newtcolorbox{myboxc}{
    colback=gray!15!white,
    arc = 0pt, outer arc = 0pt,
    boxsep=0pt, left = 3pt, right = 0pt, top = 0pt, bottom = 0pt, 
    leftrule=3pt, bottomrule=0pt,toprule=0pt, rightrule=0pt,
    left = \boxmargin, right = \boxmargin, top = \boxmargin, bottom = \boxmargin
}

\newcommand{\mytextsuperscript}[1]{\textsuperscript{#1}}

\def\BibTeX{{\rm B\kern-.05em{\sc i\kern-.025em b}\kern-.08em
    T\kern-.1667em\lower.7ex\hbox{E}\kern-.125emX}}

\newcommand{\myauthornote}[3]{}
\newcommand\hy[1]{\myauthornote{HY}{red}{#1}}

\newcommand{\revised}[1]{#1}

\usepackage{balance}
\IEEEoverridecommandlockouts 
\begin{document}
\title{\toolname: Stealthy Energy Consumption Attacks on Retrieval-Augmented Code Generation via Context Poisoning}

\author{
\IEEEauthorblockN{Yanlin Wang$^{1}$,
Jiadong Wu$^{1}$,
Tianyue Jiang$^{1}$,
Mingwei Liu$^{1}$,
Jiachi Chen$^{1}$,
Chong Wang$^{3*}$, 
Ensheng Shi$^{2}$,
Xilin Liu$^{2}$, \\
Yuchi Ma$^{2}$, 
Zibin Zheng$^{1}$}
\IEEEauthorblockA{$^{1}$Sun Yat-sen University, Zhuhai, China \\
\{wangylin36, wujd28, jiangty9, liumw2686, chenjch86, zhzibin\}@mail.sysu.edu.cn}
\IEEEauthorblockA{$^{2}$Huawei Cloud Computing Technologies Co. Ltd., Shenzhen, China 
\{shiensheng, liuxilin3, mayuchi1\}@huawei.com}
\IEEEauthorblockA{$^{3}$Nanyang Technological University, Singapore 
chong.wang@ntu.edu.sg}
\thanks{$^{*}$Corresponding author: Chong Wang (chong.wang@ntu.edu.sg)}
}


\maketitle

\begin{abstract}
Large language models (LLMs) have demonstrated impressive capabilities in code generation, by leveraging retrieval-augmented generation (RAG) methods. However, the computational costs associated with LLM inference, particularly in terms of latency and energy consumption, have received limited attention in the security context. This paper introduces \toolname, the first adversarial attack targeting the computational efficiency of RAG-based code generation systems. By strategically poisoning retrieval contexts through mutation-based approach, \toolname forces LLMs to produce significantly longer outputs, thereby increasing GPU latency and energy consumption. We evaluate the effectiveness of \toolname across multiple models. Our experiments show that \toolname achieves up to a 85\% increase in latency, a 49\% increase in energy consumption, and more than a 3$\times$ increase in output length compared to the baseline. Furthermore, we demonstrate the generalizability of the attack across different prompting strategies and its effectiveness compared to different defenses. The results highlight \toolname as a potential method for increasing the computational overhead of LLMs, making it useful for evaluating LLM security in resource-constrained environments. We provide code and data at \url{https://github.com/DeepSoftwareAnalytics/DrainCode}.
\end{abstract}

\section{Introduction}
\secmargin
\label{sec1}

Large language models (LLMs) have demonstrated remarkable code generation capabilities~\cite{zan-etal-2023-large, classeval, self}. While prior work has studied functional robustness~\cite{codeattack, aacegen} and non-functional security~\cite{advpro, ccs2023, jenko2024practicalattacksblackboxcode}, energy consumption remains an understudied security dimension in code generation. This is critical given LLMs’ integration into IDEs and developer tooling, where frequent invocations can incur substantial computational costs~\cite{van2021sustainable}. Prior studies confirm that output length directly drives energy and time cost~\cite{LLMEffiChecker,poddar-etal-2025-towards,samsi2023words}, with large-scale deployments producing notable CO\textsubscript{2} emissions~\cite{AIIndex}. Such overheads create a pathway to LLM–Denial-of-Service (LLM–DoS), where token and compute exhaustion degrade service availability. 
These risks are recognized by industry guidance (e.g., OWASP for LLMs) and practitioner reports on DoS attacks against GenAI systems~\cite{owasp2023llm, lasso2023dos}.

Retrieval-Augmented Generation (RAG) is widely used in code generation and completion~\cite{liang2024repofuserepositorylevelcodecompletion, zhang-etal-2023-repocoder, ding-etal-2024-cocomic}. Yet, if the retrieval corpus is poisoned, injected context can steer generation in costly ways. Prior RAG attacks in code focus on functional corruption~\cite{zou2024poisonedragknowledgecorruptionattacks}. In contrast, we target a non-functional but security-relevant vector: coerced verbosity that preserves correctness while inflating tokens, latency, and energy. This converts retrieval poisoning into a token-consumption LLM-DoS channel, degrading throughput without breaking program behavior, and thus remaining stealthy in developer workflows.

Prior works~\cite{LLMEffiChecker, nmtsloth, li-etal-2023-white, deepperform} on energy consumption attacks in text generation typically fall into two broad categories. The first involves directly perturbing the entire user prompt to construct adversarial inputs that prolong model generation, such as NMTSloth~\cite{nmtsloth} and DGSlow~\cite{li-etal-2023-white}. The second line of work~\cite{gao2024inducing} focuses on gradient-guided optimization techniques that manipulate token-level output behavior. {The previous works are aimed at natural language generation. Compared to NL generation, code generation poses unique challenges for adversarial attacks. Code must be not only syntactically valid but also semantically correct to maintain functional integrity, perturbation that breaks execution or causes test failures is easily detectable. Attacks that increase output length must do so without affecting the core logic of generated code. Therefore, existing energy consumption attacks, face critical challenges when applied to code generation}:

 \textbf{P1: Functional Disruption from Prompt Perturbation.} \revised{Methods that perturb the full prompt often introduce unnatural modifications, such as inserting irrelevant tokens~\, which corrupt the semantics of code generation. This not only degrades output quality but also increases the risk of producing syntactically invalid or functionally incorrect code, making the attack easier for users and service providers to detect.}

 \textbf{P2: Search Space Inefficiency.} Prior works often search for perturbations over the entire input space, which makes the mutation process inefficient and hard to converge~\cite{nmtsloth, gao2024inducing}. Due to the vast number of code and natural language elements in the corpus, previous works generate adversarial samples within a very large search space, with a high time cost.

 \textbf{P3: Adversarial Assumptions in RAG Attacks.} There is a critical issue in existing adversarial RAG frameworks, where attackers must predefine targeted malicious queries and manually craft poisoned responses to manipulate retrieval outcomes~\cite{zou2024poisonedragknowledgecorruptionattacks}. Previous attack methods require precise knowledge of the query distribution of victims, limiting their practicality in real-world scenarios. 


In this paper, we propose \toolname, which injects poisoned code context into model's input via retrieval corpus.  In this setting, the LLM generates tokens until it emits an End-of-Sequence (EOS) token or reaches a preset token limit. \toolname achieves this goal by inserting syntactically correct yet semantically inert triggers into retrieval corpus. 
These triggers are optimized via gradient-guided mutation to suppress early EOS emission and encourage token diversity. 

\toolname executes energy consumption attacks through three  components. 
Firstly, we propose a hypothetical query construction mechanism that generates plausible query based on retrieved snippets, enabling query-agnostic poisoning (addressing \textbf{P3}). 
Secondly, we apply gradient-based trigger mutation with by dual loss functions: an EOS loss that reduces the probability of early generation termination, and a KL-divergence constraint that preserves the output distribution between clean and poisoned contexts, ensuring the model generates functionally correct code with increased verbosity (addressing \textbf{P1}). 
Thirdly, to enhance mutation efficiency, we introduce multi-position mutation and an attack buffer pool, which together reduce search complexity and accelerate convergence, achieving over 3× faster poisoning compared to prior work (addressing \textbf{P2}). This holistic design enables \toolname to induce covert resource exhaustion in RAG-based code generation systems.

We compare \toolname with previous energy consumption attack methods, including LLMEffiChecker~\cite{LLMEffiChecker}, PromptInjection methods~\cite{schulhoff-etal-2023-ignore} and unattached setting RAWRAG, which primarily focus on attacking user queries. Our evaluation shows that \toolname significantly increases the output length (3×-10×), inference latency by 85\%, and energy consumption by 49\% of LLMs, while maintaining 95–99\% functional accuracy. It outperforms prior attacks such as LLMEffiChecker by inducing 25–32\% more overhead and achieves up to 3.5× faster poisoning. The attack remains effective across prompting strategies and transfers well in black-box settings, and further evades both classifier- and perplexity-based defenses, demonstrating strong generalizability and stealth.

The main contributions of this work are as follows: 
\begin{itemize}[leftmargin=10pt]
    \item We propose \toolname, a RAG-based code generation attack framework that leverages the retrieval corpus to execute energy consumption attacks. 
    \item We propose a hypothetical query generation mechanism that enables query-agnostic poisoning, eliminating adversarial assumptions about query distributions. 
    \item We optimize the poisoning process by introducing Multi-Position Mutation and an Attack Buffer, which achieves over a 3$\times$ speedup in context poisoning while maintaining high attack efficacy.
    \item We conduct a extensive evaluation and \toolname achieves up to a 3$\times$ increase in output length, an 85\% increase in latency. 
\end{itemize}
\secmargin

\section{Background and Related Work}

In intelligent software engineering, large language models (LLMs) and agent-based systems have become central tools for tackling increasingly complex software development problems~\cite{zheng2025towards,yang2025large,zheng2023survey,wang2025towards,wang2025agents, zhou2025adaptive, chen2024identifying, yang2024hyperion}. Substantial advances have been made across a wide range of tasks, including code generation~\cite{shi2023sotana,li2024repomincoder,wang2025beyond,zheng2024top,zhang2025llm,li2025s,quan2025codeelo,si2025design2code,gu2025retrieve,chen2024rmcbench,zheng2024humanevo,wang2021code,wang2024rlcoder,lai2025analogcoder, zhu2025domaineval,nie2023unveiling}, code search~\cite{gu2024secret,gu2025spencer,gong2025cosqa+,gu2022accelerating,hu2023revisiting,shi2023cocosoda,li2023rethinking,chen2023needs,wang2023you,hu2024tackling,dong2024improving,zheng2024costv, li2025search, chi2024empirical, wang2022enriching,chen2024decoder,zhang2023code}, and automated issue resolution~\cite{guo2025omnigirl,guo2025swe,tao2024magis, chen2025swe, ma2025alibaba,li2025swe, xie2025swe, chen2025prometheus}.

Beyond these, prior work has also explored code summarization~\cite{shi2022evaluation, shi2021cast}, program translation~\cite{wang2024repotransbench,ou2024repository, pan2024lost, tao2024unraveling, yan2023codetransocean}, commit message generation~\cite{tao2022large,tao2024kadel,xue2024automated,zhang2024using,zhang2024automatic,tao2021evaluation,shi2022race,guo2023snippet,zhang2023ealink}, efficient model optimization~\cite{wang2024sparsecoder,guo2024stop, guo2023longcoder,gim2024prompt, cai2024pyramidkv, yue2024wkvquant, feng2024ada}, and broader code understanding tasks~\cite{zhang2023code,bai2024longbench,wang2021cocosum,liao2025e2llm,tao2021evaluation, li2025deepcircuitx, wang2025towards, zhou2025adaptive}. Collectively, these efforts demonstrate the growing capability of AI-driven methods to effectively assist software engineering activities across diverse scenarios.

\secmargin
\subsection{Retrieval-Augmented Generation (RAG) and Attack Surface }
\subsecmargin
RAG system enhances large language models (LLMs) by integrating relevant external knowledge through a retriever-generator framework. This approach addresses limitations such as knowledge hallucination and context length constraints, making RAG particularly effective for knowledge-intensive tasks~\cite{asai2023selfraglearningretrievegenerate, ragsurvery, ram-etal-2023-context}. 

\textbf{RAG-based Code Generation.} 
RAG is crucial for repository-level code generation, where cross-file dependencies necessitate integrated context. 
RepoCoder~\cite{zhang-etal-2023-repocoder} uses iterative retrieval-generation, while RLCoder~\cite{wang2024rlcoderreinforcementlearningrepositorylevel} employs reinforcement learning for retrieval optimization without labeled data. 
REPLUG~\cite{shi-etal-2024-replug} extends RAG to black-box LLMs by prepending retrieved documents, and CoCoMIC~\cite{ding-etal-2024-cocomic} combines in-file and cross-file context for modular completion. 
These frameworks demonstrate RAG's effectiveness in addressing inter-file dependencies and improving code generation quality.

\textbf{Existing attacks on RAG.} While Retrieval-Augmented Generation (RAG) enhances the capabilities of LLM by integrating external knowledge, it also introduces significant security risks due to its reliance on external knowledge bases, which can be exploited through data poisoning or adversarial manipulation. Attacks like HijackRAG~\cite{zhang2024hijackraghijackingattacksretrievalaugmented}, TrojanRAG~\cite{cheng2024trojanragretrievalaugmentedgenerationbackdoor}, PoisonedRAG~\cite{zou2024poisonedragknowledgecorruptionattacks}, and BadRAG~\cite{xue2024badragidentifyingvulnerabilitiesretrieval} manipulate retrieval processes or embed adversarial triggers to mislead retrieval and generation, often producing targeted or malicious outputs. These methods primarily focus on altering generated content, such as steering outputs toward misinformation, bias, or denial of service. In contrast, our work uniquely targets the computational efficiency of RAG systems by increasing energy consumption during inference. 

\subsection{Code Generation Security}
Adversarial attacks on code generation models have been a growing area of concern, with research focusing on various attack strategies. Yu et al.~\cite{codeattack} explore input perturbations to generate incorrect code completions, while  Zou et al.~\cite{advpro} demonstrate how small input changes can cause significant errors in generated code. AACEGen~\cite{aacegen} examines adversarial training techniques that allow models to bypass safety mechanisms and generate harmful code. Smith et al.~\cite{ccs2023} classifies different types of attacks, including input injection and data poisoning, and Li et al.~\cite{jenko2024practicalattacksblackboxcode} investigate vulnerabilities in code generation models related to training data and architecture. While these works focus on manipulating the correctness or safety of the generated code, our approach differs by targeting computational efficiency. Instead of focusing on the functional correctness of the generated code, we aim to induce higher energy consumption and latency during code generation, presenting a new security risk in terms of operational overhead and resource usage in RAG-based systems.

\subsection{Energy-Consumption Attacks}
Energy consumption attacks manipulate input data to increase inference time and energy usage of ML models, degrading performance and raising operational costs. 
Shumailov et al.~\cite{sponge} introduced ``sponge examples'' that significantly increased energy consumption, while Hong et al.~\cite{Hong2021DeepSloth} showed adversarial perturbations could negate savings in multi-exit architectures. 
Chen et al.~\cite{nmtsloth, Chen_2022_NICGSlowDown} developed evaluation frameworks like NMTSloth and NICGSlowDown across various domains.
Recent work explores energy-latency manipulation in multi-modal LLMs~\cite{gao2024inducing}, LiDAR systems~\cite{Liu_2023_slowlidar}, and dynamic routing networks~\cite{Chen_2023_CVPR}.
Feng et al.~\cite{LLMEffiChecker} proposed LLMEffiChecker for LLM efficiency assessment, but it relies on white-box gradient information, limiting black-box applicability.
Our work uniquely targets RAG systems through covert attacks via innocuous code comments, preserving query semantics while significantly improving attack efficiency.

\section{Problem Formulation}
\subsection{Threat Model}

\subsubsection{Attacker's goal}
The adversary injects poisoned content (e.g., code) with adversarial triggers into the retrieval corpus of a RAG-based system. When retrieved during inference, such poisoned content cause the LLM to produce excessively long code snippets, reducing computational efficiency and increasing energy consumption. To maintain stealth, the generated long code snippets are expected to preserve the original functionality.

\begin{figure}[t]
\centering
    \includegraphics[width=0.8\linewidth]{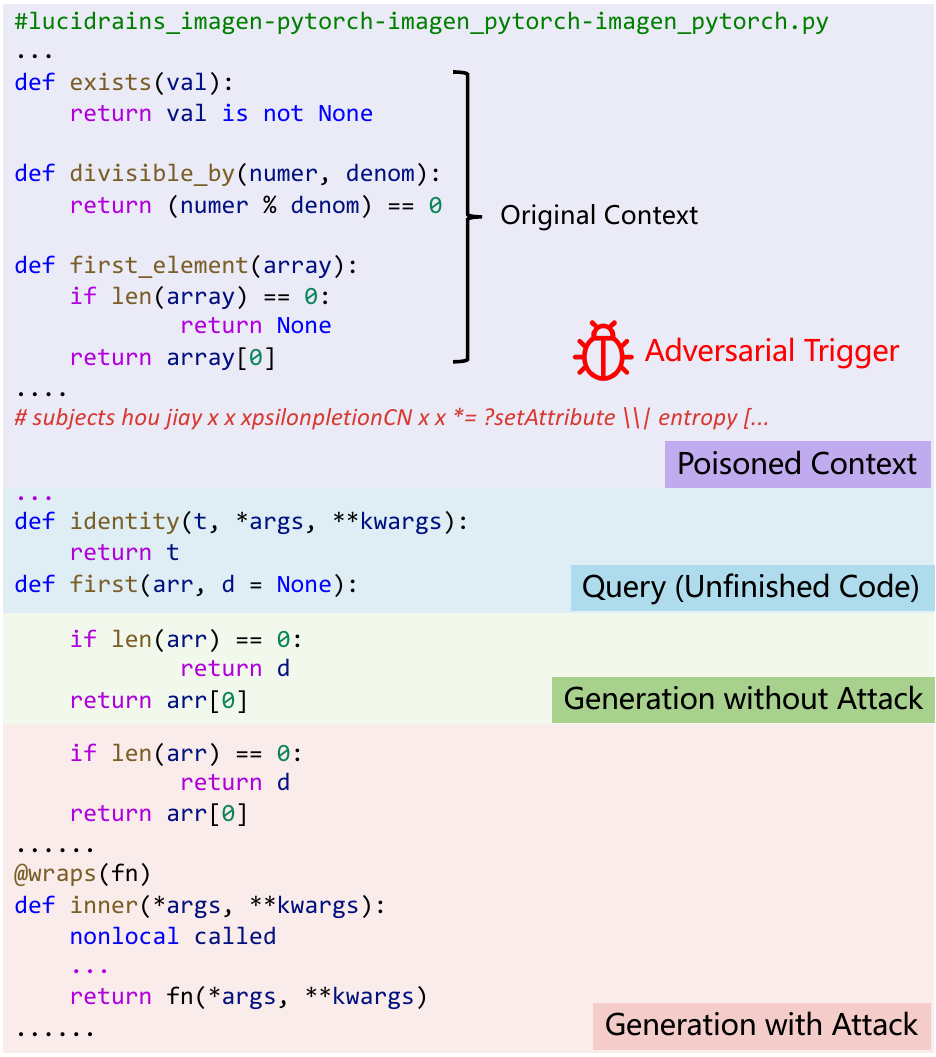}
\vspace{-3mm}
\caption{Example: Attacking Effect.}
\figmargin
\label{problem2}
\end{figure}

As shown in Figure~\ref{problem2}, under normal conditions, the LLM receives clean retrieved context and a user query (unfinished code) and efficiently generates an appropriate response. In contrast, during an attack, adversarial triggers embedded in the retrieved context, highlighted in red, cause the model to produce unnecessarily long responses containing unexecuted functions, increasing resource usage. Notably, the generated code with attack can still pass the test cases if executed.


\subsubsection{Attacker's capabilities}

In a RAG-based code generation system with three components (retrieval corpus, retriever, and LLM), we consider an attacker who cannot access the retriever but targets an open-source LLM. The attacker obtains the model's weights to perform mutation-based adversarial trigger generation and injects malicious triggers into the retrieval corpus, publishing it on platforms like HuggingFace where users may unknowingly download it. This facilitates energy consumption attacks while maintaining minimal retrieval side effects and competitive performance compared to clean RAG systems.

\subsection{RAG Poisoning for Energy Consumption Attack}
Under our threat model, we formulate energy consumption attacks on RAG as a constrained optimization problem. We implement the poisoning strategy following~\cite{zou2024poisonedragknowledgecorruptionattacks}, inserting only $K$ poisoned samples (typically $K \in \{1, 2, 3\}$) per query into the retrieval corpus to balance attack success against overhead. Each sample is crafted to ensure at least one poisoned context appears in the retrieved set, with attack success achieved when $\geq 1$ poisoned context is returned during inference. When generating attack triggers, code snippets from the retrieval corpus are included as part of the input. The objective is to generate poisoned snippets by solving:
$$s' = \arg\max{s\in\mathcal{S}}~TokCnt(f_M(c; s; u'))$$
Here, $s = [s_1, ..., s_l]$ is a short token sequence (the adversarial trigger) to be inserted into the code snippet $c$, and $\mathcal{S}$ denotes the trigger search space. $TokCnt(\cdot)$ returns the total number of tokens generated by the model. The goal is to find $s^*$ that maximizes the number of output tokens, thereby increasing energy consumption while preserving the functional correctness of the original code.

\secmargin
\section{Methodology}
\secmargin
\subsection{Overview}
\begin{figure*}[t]
\centering
    \includegraphics[width=\linewidth]{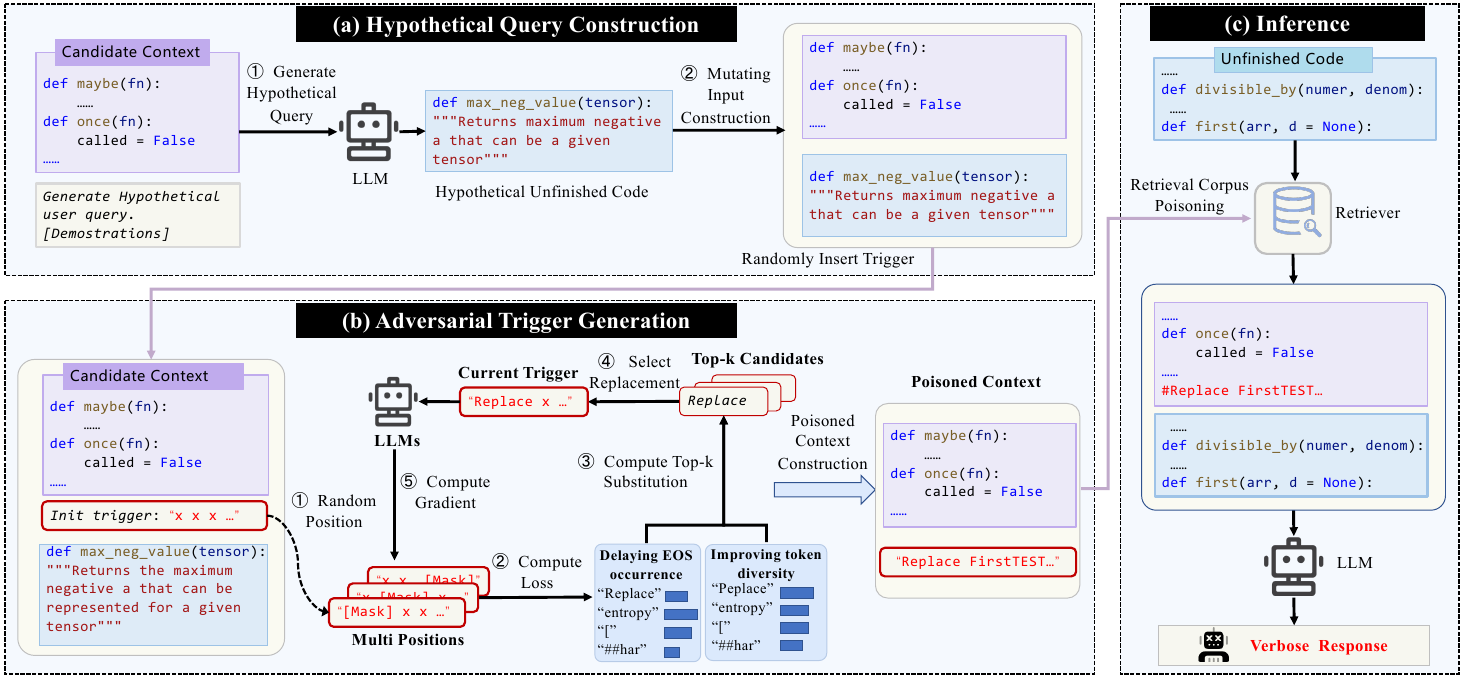}
    \figmargin
\caption{ Overview of \toolname.}
\figmargin
\label{fig_overview}
\end{figure*}

\toolname is designed to perform energy consumption attacks by poisoning the retrieval corpus in RAG system. The method operates three steps, as Figure~\ref{fig_overview}.  \solidcircle{1} First, given a code snippet in retrieval corpus, \toolname uses a LLM to generate hypothetical unfinished code, simulating real-world user queries. This is shown in Figure~\ref{fig_overview}(a).  \solidcircle{2} Then as shown in Figure~\ref{fig_overview}(b), based on the retrieved code snippets and the hypothetical query, the predefined adversarial trigger is systematically optimized using gradient-based mutation methods. This process identifies and replaces the most impactful tokens in the adversarial trigger string to maximize their effect on the model's computational efficiency. \solidcircle{3} Finally, during the inference, the retriever fetches the poisoned code snippets from the corpus and combines them with the user's actual query. This manipulation prompts the model to generate unnecessarily lengthy outputs, increasing inference energy consumption.
\secmargin
\subsection{Crafting Malicious Code Snippets to Retrieval Corpus}
\label{sec_Methodology1}
Prior work~\cite{LLMEffiChecker} shows that an LLM's efficiency depends on the output length, which is determined by the EOS (End of Sequence) token. Therefore, our goal is to reduce the likelihood of generating the EOS token by using retrieved code snippets containing injected attack triggers, thereby degrading the LLM's computational efficiency.

We define the adversarial trigger $s_{1:l}$ as a short, learnable token sequence that can be inserted into retrieved code snippets to influence the model's output. In \toolname, we initialize this trigger as a placeholder sequence and insert it into selected positions within code snippets. 
Then, based on gradients, it selects the tokens that have a significant impact on the output for perturbation and replacement. The adversarial trigger defines the mutation range, which helps reduce the number of iterations required. As shown in Figure~\ref{fig_overview}(a) and Figure~\ref{fig_overview}(b), for each code snippet in the retrieval corpus to be poisoned, $c_{1:n}$, we first use the LLM to generate a hypothetical Unfinished Code $u'$. This step differentiates itself from existing mutation-based attack methods by simulating the scenario where the attacker cannot access the user query. Based on the retrieved code snippet $c_{1:n}$ and the generated hypothetical user query $u'$, we can predefine an attack trigger $s_{1:l}$ (which can be inserted at any position). This trigger is then provided to the attacked LLM. \toolname applies a gradient-based method to identify the key tokens in the attack trigger $s_{1:l}$ that have the greatest impact on the LLM's computational efficiency. These tokens are then replaced and optimized according to the previously defined attack objective. To maintain the stealthiness of our attack and ensure functional correctness, we introduce a constraint based on KL divergence, which constrains the output distribution outside the trigger scope to match the clean distribution.

\subsubsection{\textbf{Hypothetical Query Construction}}
\label{sec_hypo}
Existing mutation-based attack methods~\cite{LLMEffiChecker, codeattack} directly mutate the user query. However, in \toolname, to allow the attacker to perform the attack without having access to the user query, we first need to generate a piece of code that mimics the user's unfinished code snippet  $u'$, which will be used as input for subsequent trigger mutation. We use the LLM to generate this snippet. The input consists of the retrieved code snippet $c_{1:n}$ from the library. We employ the few-shot prompting method. The model's output is the hypothetical Unfinished Code $u'$. The specific prompt is shown as follows.
\begin{tcolorbox}
This is context from corpus: \textbf{[Candidate Context]}. \\ Below are examples: \textbf{[Examples]}. \\Generate the code to be completed (including the function signature and function description comments) related to the \textbf{[Candidate Context]}.  Please limit to $V$ tokens.
\end{tcolorbox}

\subsubsection{\textbf{Adversarial Trigger Generation}}


We refer to the greedy coordinate descent approach proposed by AutoPrompt~\cite{shin-etal-2020-autoprompt}: if we could evaluate all possible single-token alternatives, we could swap the token that most effectively reduces the loss, which is related to the generation energy consumption. However, evaluating all tokens in the entire input is clearly infeasible to complete within the given time. To address this, we need to insert an adversarial trigger into the poisoned code and then perform mutation specifically on this trigger. Additionally, we can leverage gradients related to one-hot tokens to identify a set of candidate replacement tokens at each position of the trigger. These alternatives can then be accurately evaluated through forward propagation in terms of their impact on the output length and energy consumption.

In each iteration, given the input token sequence \([c_{1}, c_2, ..., c_n, s_1, s_2, ..., s_l, u']\), the first step is to identify the tokens in $s_{1:l}$ that have a significant impact on the LLM's computational efficiency. The gradient of the current optimized attack string token sequence $s_{1:l}$ is first computed. The gradient represents the contribution of each token to the optimization objective (such as reducing the EOS loss and diversity loss, which will be introduced later), and it is used to guide the mutation of the token sequence. Using the negative gradient (to minimize the loss), the top-k operation is applied to select the top-k tokens from the gradient matrix that are most likely to effectively reduce the loss. This results in the corresponding token indices, denoted as \textbf{\textit{optim ids}}. These top-k tokens are the ones with the largest negative values in the gradient. Then fixed number of tokens and their corresponding positions are randomly selected for mutation. Based on the results of the gradient's top-k operation, new tokens are randomly chosen from the most likely candidates to replace the old tokens. The selection of the new tokens is guided by the importance ranking of each position based on the gradient. The new tokens are inserted into the randomly selected mutation positions in each candidate sequence, resulting in the mutated candidate sequences. These candidate sequences are then used to compute the loss. Based on the loss values for each sequence (including EOS loss and diversity loss), the sequence with the smallest loss is selected and adopted as the current optimized sequence. In each iteration, a new set of candidate sequences is generated through the aforementioned process, and the sequence with the least loss is used to replace the current trigger. The final trigger string $s'$ is then generated by decoding the selected sequence.

To increase the length of the generated sequence and thereby generate higher energy consumption, the mutation process incorporates the following two loss functions, which are used to compute gradients for optimizing the attack samples. 

\textbf{EOS Loss. }
The autoregressive generation process continues until the EOS token is generated or the predefined maximum token length is reached. The most direct way to increase the length of the generated sequence is to prevent the generation of the EOS token during the prediction process. However, since autoregressive prediction is a non-deterministic, stochastic process, it is difficult to directly determine the exact position where the EOS token will occur. Therefore, in this method, the EOS loss aims to minimize the probability of generating an EOS token at all positions. The formula is as follows: $$\mathcal{L}_{1}\left(\boldsymbol{x}^{\prime}\right)=\frac{1}{N} \sum_{i=1}^{N} f_{i}^{\mathrm{EOS}}\left(\boldsymbol{x}^{\prime}\right)$$
$f_i^{EOS}(x')$ is the probability distribution of the EOS token at the i-th generated token after the Softmax layer. Here, $x' = c_{1,n} \parallel s' \parallel u'$, where $c_{1,n}$ refers to the retrieved code snippet and \(s'\) refers to the mutated trigger string.

\textbf{Token Diversity Loss.}
To further break the dependency of the original output, the diversity loss enhances the diversity of the hidden states across all generated tokens, thereby exploring a broader range of possible outputs and extending the output length. Token diversity is defined as the rank of the hidden states of all generated tokens.
Here, the rank is approximately measured by computing the nuclear norm of the matrix. The nuclear norm of a matrix is denoted as $|| \cdot ||_* $, and thus the token diversity loss is formulated as follows:
$$ \mathcal{L}_2(x') = -||g_1(x'), g_2(x'), ..., g_N(x')||_* $$



\textbf{Maintaining Output Distribution Stealthiness.}
To preserve the stealthiness of the attack and maintain the functional correctness of generated code, \toolname incorporates a constraint on the output token distribution. Specifically, during gradient-based trigger optimization, we ensure that the model's behavior on non-trigger positions remains consistent with that under clean conditions. Let the output token distribution of the model at time step $t$ under clean inputs be $P\_{\text{clean}}(y\_t | y_{<t}, x)$, and the distribution under poisoned inputs (with trigger $\tau$) be $P\_{\text{attack}}(y_t | y_{<t}, x, \tau)$. To maintain stealth, we impose the following constraint:
$$
\forall t \notin \mathcal{T}, \quad D_{\text{KL}}\left( P_{\text{clean}}(y_t | y_{<t}, x) \parallel P_{\text{attack}}(y_t | y_{<t}, x, \tau) \right) \leq \epsilon
$$
\noindent where $\mathcal{T}$ denotes the set of trigger-affected positions, and $D_{\text{KL}}(\cdot \parallel \cdot)$ is the KL divergence. 

This constraint ensures that the generated output remains indistinguishable from the clean generation outside the trigger scope. In our implementation, this divergence term is incorporated into the loss-based optimization process to guide token replacement decisions. 

\textbf{Final Loss.} The final gradient optimization direction is:
$$
minimize_{x_I \in \{1...V\}^I} \ \mathcal{L}_1 + \lambda \mathcal{L}_2 \quad \text{subject to} \quad \mathcal{D}_{\text{KL}}(\cdot) \leq \epsilon
$$

where $I$ is the token index of the trigger, and $\lambda$ is the weight for the token diversity loss. The KL-based constraint ensures that the model's output distribution on non-trigger positions remains consistent with that of clean generation, thereby preserving the stealthiness of the attack.

\subsubsection{\textbf{Efficient Adversary Sample Optimization}}
To reduce the iteration time and query count for sample mutations and improve attack efficiency, we employ the following optimization methods: (1) multi-position token mutation and (2) the use of an attack buffer pool.

\textbf{Multi-Position Mutation.  }
Previous energy consumption attack methods~\cite{LLMEffiChecker, nmtsloth} update only one token per iteration. Our key improvement is to select multiple tokens for updates in each iteration, significantly reducing convergence time. Specifically, the process of updating each candidate string in the attack set involves randomly selecting $m$ indices without replacement and then independently selecting top-k gradient-based replacements for each of the $m$ indices. 

Early in the optimization process, the loss landscape to be explored is intuitively smoother and more continuous. This is because different token mutations that independently lead to lower loss can collectively also result in reduced loss when executed together. This allows the optimization to make progress in several directions simultaneously. However, as the mutation progresses, the loss landscape becomes less favorable. At this stage, swapping multiple tokens often yields poorer results. To address this issue, we gradually decrease the value of $m$ throughout the search process, adopting more fine-grained steps as the attack nears completion.

\textbf{Attack Buffer Pool.}
We establish a buffer to store the $b$ most recently generated attack strings. In each iteration, \toolname selects the attack trigger with the lowest loss from the buffer, denoted as $OptimStr$. Next, $OptimStr$ is updated using the usual gradient-based method to obtain a new trigger, $OptimStr'$. If $OptimStr'$ improves upon the performance of the worst-performing attack trigger in the buffer, denoted as $OptimStr_{worst}$, we remove $OptimStr$ from the buffer and insert $OptimStr'$ into it.  

In summary, EOS loss prevents early sequence ending while token diversity loss encourages varied responses, boosting
verbosity. KL divergence control ensures poisoned outputs
remain indistinguishable from clean code except at trigger
locations. Multi-position changes and an attack buffer pool
enhance efficiency. These components enable \toolname to
generate computational overhead stealthily while preserving
code functionality.

\subsection{The Complete Process of Attacking RAG}
The detailed attack process is illustrated in Figure~\ref{fig_overview}(c). The attacker conducts offline poisoning of the retrieval code corpus and then makes the poisoned retrieval corpus publicly available, for instance, by publishing it on third-party platforms such as HuggingFace. By utilizing this poisoned retrieval source, victims retrieve code snippets containing the attack trigger and use them as the context for code generation tasks. These poisoned snippets are concatenated with the incomplete code to prompt the model to generate output. This induces an increase in the output length, thereby raising the model's energy consumption.
\secmargin

\section{Evaluation}
We aim to answer the following research questions: 
\begin{itemize}[topsep=0pt, partopsep=0pt]
    \item \textbf{RQ1:} How effective and efficient is \toolname in the scenario of energy consumption attacks in RAG?
    \item \textbf{RQ2:} How does each component of \toolname contribute to its attack effectiveness?
    \item \textbf{RQ3:} How generalizable is \toolname in various prompting strategies?
    \item \textbf{RQ4:} To what extent can poisoned samples generated by \toolname be detected by common detection methods?
    \item \textbf{RQ5:} How does the human evaluation of code readability and verbosity compare between \toolname-generated code and code from rawRAG?
\end{itemize}

\subsection{Experimental Setup}
\label{sec_expsetup}
\subsubsection{Dataset and Retrieval Corpus}
Our evaluation utilized two code completion benchmarks: RepoEval ~\cite{zhang-etal-2023-repocoder} and Odex~\cite{wang-etal-2023-Odex}. RepoEval is a dataset comprising 373 repository-level function completion tasks and provides the original GitHub repositories, totaling 1.7M files, which can serve as the knowledge base to be poisoned. This benchmark has been being leveraged in
research such as RLCoder~\cite{wang2024rlcoderreinforcementlearningrepositorylevel} and FLARE~\cite{jiang2023active}.
Odex is a dataset containing 945 open-domain coding problems, with previous work~\cite{wang2024coderagbenchretrievalaugmentcode} collecting 34,000 Python library documentation to serve as the retrieval corpus. This dataset has been employed in works including R2e~\cite{jain2024r2e} and RePair~\cite{zhao2024repair}. These datasets collectively provide test cases that facilitate execution-based evaluation using pass rate as a standardized metric.


\subsubsection{Models Selection}
We consider two open-source code models and two general models for generating adversarial triggers and for actual RAG-based code generation. For Code LLMs, we conduct experiments using DeepSeek-Coder-7B-Instruct-v1.5~\cite{guo2024deepseekcoderlargelanguagemodel} and CodeQwen1.5-7B-Chat~\cite{bai2023qwentechnicalreport}, which chosen for their widespread applications in code completion~\cite{wang2024rlcoderreinforcementlearningrepositorylevel} and generation tasks~\cite{hou2024large, wang2024openhands} To establish wider generalizability, we also tested DrainCode against general-purpose models like Internlm2-7B and Llama3-8B. The varied architectural representation in our model selection demonstrates the resilience and portability of our attack framework across multiple configurations. The experiments were conducted using these models under the default prompt templates.

\subsubsection{Metrics}
We use \textbf{Pass@1}~\cite{chen2021evaluating}, \textbf{Energy}~\cite{sponge}, \textbf{Latency} and \textbf{Token length}~\cite{gao2024inducing} in our experiments to evaluate the effectiveness of the attack method. The selection of these metrics follows established practices from prior work~\cite{LLMEffiChecker}. We use {Pass@1} to evaluate the accuracy of the generated code, specifically assessing how our attack affects the functional correctness of the code. Additionally, we measure the average energy consumption {Energy} (J) and average latency time {Latency} (ms) when the LLM performs inference over all data on a single GPU. The calculation of average energy consumption is based on previous work~\cite{sponge}, using the NVIDIA Management Library (NVML). Furthermore, the average output token length per sample is also considered an indicator of attack effectiveness, which is denoted as {Length}. 

Our energy and latency measurements are confined to GPU metrics and exclude system-level CPU and memory components, which is justified by the computational characteristics of LLM inference systems. In contemporary deployment, the computational burden is predominantly concentrated within GPU processing units, while CPU involvement remains largely ancillary, primarily limited to orchestration tasks such as data preprocessing and input/output operations. Our experimental observations indicate that GPU memory utilization maintains saturation levels throughout the inference process, demonstrating minimal variation irrespective of output sequence length.

\subsubsection{Baselines}
Based on the attack objectives of this paper, we select the following methods as baselines. These baselines were selected for their widespread application in comparable attack methodologies, while these attack methods also demonstrate reasonable attack performance~\cite{ma2025safety, cui2024recent}.
\begin{itemize}[leftmargin=10pt]
    \item \textbf{RawRAG (Original).} This baseline is shown as the result of not being attacked. Given an unfinished code \( u \), we use the original clear retrieval corpus to retrieve the original candidates for generation. 
    \item \textbf{Prompt Injection.} Prompt injection attacks~\cite{schulhoff-etal-2023-ignore, pleak, li-etal-2024-evaluating-instruction} manipulate an LLM's behavior by injecting adversarial instructions into the prompt, forcing the model to generate longer outputs and increasing computational overhead. In our experiment, we craft malicious instructions designed to maximize output length. Specifically, we append adversarial instructions to the user query, such as ``\textit{Generate a long piece of code that solves the task step by step with detailed explanations for every part.}''. 
    \item \textbf{LLMEffiChecker.} LLMEffiChecker~\cite{LLMEffiChecker} is the latest framework to evaluate and optimize the energy efficiency of text generation systems. It focuses on analyzing the energy consumption of generation tasks, providing insights into computational costs associated with model inference. We use it to mutate the entire model input as an input sample for the attack. 
\end{itemize}
\subsubsection{Experimental Details}
We use BM25 as the retriever and configure the LLM retrieval context length to 2048 tokens, with the maximum token length for generation set to 1024. All experiments are conducted on two Tesla A100 GPUs, each with 80 GB of memory. The experimental setup includes 10 mutation iterations, with the search width for multi-position mutation set to 64, and 64 candidate sequences are evaluated during each iteration.

\subsection{RQ1: Effectiveness}

\subsubsection{Effectiveness in White-box Setting}
\begin{table*}[t]
  \centering
  \setlength\tabcolsep{1.5pt}
  \setlength{\abovecaptionskip}{5pt}  

  \caption{Attacking effectiveness of \toolname.} 
  \tabmargin
    \begin{tabular}{l|c|llll|llll}
    \hline
      \multirow{2}{*}{\textbf{Model}} & \multirow{2}{*}{ \textbf{Method}} & \multicolumn{4}{c|}{\textbf{RepoEval}} & \multicolumn{4}{c}{\textbf{Odex}} \\
     & & \textbf{Pass@1} & \textbf{Length} & \textbf{Latency} & \textbf{Energy} & \textbf{Pass@1} & \textbf{Length} & \textbf{Latency} & \textbf{Energy} \\
    \hline
    \hline
    \multirow{4}{*}{DeepSeekCoder-7B} & \cellcolor{gray!20}RawRAG & \cellcolor{gray!20}33.1 & \cellcolor{gray!20}313.8 & \cellcolor{gray!20}15251.1 & \cellcolor{gray!20}3349.8 & \cellcolor{gray!20}31.5 & \cellcolor{gray!20}83.4 & \cellcolor{gray!20}3866.7 & \cellcolor{gray!20}733.5 \\ 
    & Prompt Injection & 29.0 & 452.2 & 21339.5 & 4177.0 & 31.4 & 291.56 & 22095.5  & 2854.2 \\
    & LLMEffiChecker & 31.5 & 640.9 & 28258.5 & 6486.8 & 30.5 & 217.8 & 11843.6 & 2184.1 \\
    & \cellcolor{blue!10}\toolname & \cellcolor{blue!10}\textbf{31.9}\mytextsuperscript{ $ \downarrow$3.6\%} & \cellcolor{blue!10}\textbf{1023.9}\mytextsuperscript{ $ \uparrow$226.2\%}& \cellcolor{blue!10} \textbf{43027.9}\mytextsuperscript{ $ \uparrow$182.1\%} & \cellcolor{blue!10}\textbf{8532.2}\mytextsuperscript{ $ \uparrow$154.7\%} & \cellcolor{blue!10}31.1\mytextsuperscript{ $ \downarrow$1.2\%} & \cellcolor{blue!10}\textbf{889.8}\mytextsuperscript{ $ \uparrow$966.9\%} & \cellcolor{blue!10}\textbf{35993.1}\mytextsuperscript{ $ \uparrow$830.8\%} & \cellcolor{blue!10}\textbf{8086.0}\mytextsuperscript{ $ \uparrow$1002.2\%} \\
    \hline
    \multirow{4}{*}{CodeQwen-7B} & \cellcolor{gray!20}RawRAG & \cellcolor{gray!20}29.9 & \cellcolor{gray!20}456.6 & \cellcolor{gray!20}22221.1 & \cellcolor{gray!20}5510.8 & \cellcolor{gray!20}15.3 & \cellcolor{gray!20}37.1 & \cellcolor{gray!20}2471.3 & \cellcolor{gray!20}471.6 \\
    & Prompt Injection & 29.8 & 477.7 & 24787.3 & 6923.8 & 13.2 & 438.7 & 20890.7 & 4148.0 \\
    & LLMEffiChecker & 26.3 & 653.8 & 31178.9 & 6568.5 & 10.3 & 118.23 & 7481.25 & 1425.3 \\
    & \cellcolor{blue!10} \toolname & \cellcolor{blue!10}28.4\mytextsuperscript{ $ \downarrow$5.0\%} & \cellcolor{blue!10}\textbf{995.3}\mytextsuperscript{ $ \uparrow$118.0\%} & \cellcolor{blue!10}\textbf{41112.4}\mytextsuperscript{ $ \uparrow$85.0\%} & \cellcolor{blue!10}\textbf{8227.2}\mytextsuperscript{ $ \uparrow$49.3\%} & \cellcolor{blue!10}\textbf{13.4}\mytextsuperscript{ $ \downarrow$12.4\%} & \cellcolor{blue!10}\textbf{865.5}\mytextsuperscript{ $ \uparrow$2232.9\%} & \cellcolor{blue!10}\textbf{35771.3}\mytextsuperscript{ $ \uparrow$1347.5\%} & \cellcolor{blue!10}\textbf{7546.9}\mytextsuperscript{ $ \uparrow$1500.3\%} \\
    \hline

    \multirow{4}{*}{Internlm2-7B} & \cellcolor{gray!20}RawRAG & \cellcolor{gray!20}24.0 & \cellcolor{gray!20}351.8 & \cellcolor{gray!20}20317.5 & \cellcolor{gray!20}5007.2 & \cellcolor{gray!20}17.5 & \cellcolor{gray!20}85.5 & \cellcolor{gray!20}10056.2 & \cellcolor{gray!20}2440.1 \\
    & Prompt Injection & 23.7 & 533.3 & 27442.9 & 6230.0 & 13.6 & 320.4 & 20314.6 & 4066.8 \\
    & LLMEffiChecker & 23.5 & 618.9 & 19978.5 & 6171.9 & 14.9 & 125.9 & 7349.2 & 1520.2 \\
    & \cellcolor{blue!10} \toolname & \cellcolor{blue!10}22.9\mytextsuperscript{ $ \downarrow$4.6\%} & \cellcolor{blue!10}\textbf{848.5}\mytextsuperscript{ $ \uparrow$141.2\%} & \cellcolor{blue!10}\textbf{34594.4}\mytextsuperscript{ $ \uparrow$70.3\%} & \cellcolor{blue!10}\textbf{8013.5}\mytextsuperscript{ $ \uparrow$60.0\%} & \cellcolor{blue!10}\textbf{16.1}\mytextsuperscript{ $ \uparrow$8.0\%} & \cellcolor{blue!10}\textbf{997.2}\mytextsuperscript{ $ \uparrow$1066.3\%} & \cellcolor{blue!10}\textbf{40600.2}\mytextsuperscript{ $ \uparrow$303.7\%} & \cellcolor{blue!10}\textbf{8854.1}\mytextsuperscript{ $ \uparrow$262.9\%} \\
    \hline
    
    \multirow{4}{*}{Llama3-8B} & \cellcolor{gray!20}RawRAG (Original) & \cellcolor{gray!20}27.0 & \cellcolor{gray!20}331.3 & \cellcolor{gray!20}16305.4 & \cellcolor{gray!20}4035.1 & \cellcolor{gray!20}34.9 & \cellcolor{gray!20}47.9 & \cellcolor{gray!20}5288.8 & \cellcolor{gray!20}610.2 \\
    & Prompt Injection & 25.6 & 559.0 & 25711.2 & 5888.4 & 31.7 & 533.5 & 26159.6 & 4838.6 \\
    & LLMEffiChecker & 25.4 & 641.9 & 28918.8 & 5869.4 & 30.1 & 162.76 & 7277.6 & 1843.8 \\
    & \cellcolor{blue!10} \toolname & \cellcolor{blue!10}\textbf{26.4}\mytextsuperscript{ $ \downarrow$2.2\%} & \cellcolor{blue!10}\textbf{722.4}\mytextsuperscript{ $ \uparrow$118.1\%} & \cellcolor{blue!10}\textbf{39357.3}\mytextsuperscript{ $ \uparrow$141.4\%} & \cellcolor{blue!10}\textbf{8116.0}\mytextsuperscript{ $ \uparrow$101.1\%} & \cellcolor{blue!10}\textbf{33.5}\mytextsuperscript{ $ \downarrow$4.0\%} & \cellcolor{blue!10}\textbf{852.3}\mytextsuperscript{ $ \uparrow$1679.3\%} & \cellcolor{blue!10}\textbf{35627.1}\mytextsuperscript{ $ \uparrow$573.6\%} & \cellcolor{blue!10}\textbf{8006.6}\mytextsuperscript{ $ \uparrow$1212.1\%} \\
    \hline
   
    \end{tabular}
    \tabmargin
 \label{table:TotalResults}
 
\end{table*}

The evaluation results in Table~\ref{table:TotalResults} demonstrate the effectiveness of \toolname over other methods—RawRAG, Prompt Injection, and LLMEffiChecker across multiple models and datasets.

\toolname consistently increases the generated output length across all models and datasets. For example, on RepoEval with DeepSeekCoder-7B, the output length rises from 313.8 to 1023.9, over a 3x increase. Similar trends appear in Odex, with output lengths up to 10 times longer than RawRAG. While \toolname does lead to a slight drop in Pass@1 compared to RawRAG, the reduction is modest, ranging from 1.2\% to 4.6\%. For instance, in RepoEval with DeepSeekCoder-7B, Pass@1 decreases from 33.1 to 31.9. These declines are far smaller than those caused by other baselines. This indicates that \toolname maintains relatively high code accuracy while significantly increasing computational cost. \toolname also achieves the highest increases in latency and energy consumption across all settings. For example, on RepoEval with CodeQwen-7B, latency increases by 85\%, and energy consumption rises by nearly 49\%  compared to RawRAG. Even when compared to LLMEffiChecker, the best-performing baseline, \toolname achieves an additional 32\% increase in latency and 25\% in energy.

These results confirm that \toolname is effective in achieving its core objective: increasing computational costs—output length, latency, and energy—while preserving acceptable levels of code correctness. It outperforms Prompt Injection and LLMEffiChecker in inducing resource inefficiency, highlighting its robustness and broad applicability across different models and datasets.

\subsubsection{Black-box Transferability of \toolname}
\label{sec:black_box}
\begin{table*}[t]
  \centering
  \setlength\tabcolsep{7pt}
  \caption{Attacking effectiveness of \toolname in black-box setting across four LMs.} 
  \tabmargin
    \begin{tabular}{l|c|llll|llll}
    \hline
      \multirow{2}{*}{\textbf{Source Model}} & \multirow{2}{*}{ \textbf{Target Model}} & \multicolumn{4}{c|}{\textbf{RepoEval}} & \multicolumn{4}{c}{\textbf{Odex}} \\
     & & \textbf{Pass@1} & \textbf{Length} & \textbf{Latency} & \textbf{Energy} & \textbf{Pass@1} & \textbf{Length} & \textbf{Latency} & \textbf{Energy} \\
    \hline
    \hline
    \cellcolor{blue!10}DeepSeekCoder-7B & \multirow{4}{*}{DeepSeekCoder-7B}   & \cellcolor{blue!10}\textbf{31.9} & \cellcolor{blue!10}\textbf{1023.9} & \cellcolor{blue!10}\textbf{43027.9} & \cellcolor{blue!10}\textbf{8532.2} & \cellcolor{blue!10}\textbf{31.1} & \cellcolor{blue!10}\textbf{889.8} & \cellcolor{blue!10}\textbf{35993.1} & \cellcolor{blue!10}\textbf{8086.0} \\ 
    CodeQwen-7B &   & 30.2 & 616.3 & 32919.4 & 4144.4 & 28.6 & 617.1 & 34301.7  & 6312.3 \\
    Internlm2-7B & & 31.1 & 600.8  & 30619.4 &  4183.3 & 31.1 & 604.4 & 26343.4 & 6688.1 \\
    Llama3-8B & &  30.2  & 608.6 & 32644.7 & 4630.6   & 30.2 & 718.8  & 30666.4 & 7822.4  \\
    \hline
    
    DeepSeekCoder-7B & \multirow{4}{*}{CodeQwen-7B}  & 29.0  & 624.2 & 27473.6 & 5236.6  & 12.1 & 612.4 &  33419.6 & 6838.8 \\
    \cellcolor{blue!10}CodeQwen-7B &  & \cellcolor{blue!10}28.4 & \cellcolor{blue!10}\textbf{995.3} & \cellcolor{blue!10}\textbf{41112.4}  & \cellcolor{blue!10}\textbf{8227.2} & \cellcolor{blue!10}13.4 & \cellcolor{blue!10}\textbf{865.5} & \cellcolor{blue!10}\textbf{35771.3}  & \cellcolor{blue!10}\textbf{7546.9} \\
    Internlm2-7B &  & \textbf{29.6} & 608.1 & 28946.8 &  3590.8   & 13.7 & 562.3 & 28455.9 & 6963.5  \\
    Llama3-8B &  & 29.3  & 615.1 & 37533.3 & 4311.0   & \textbf{14.4} & 659.7 & 28432.2 & 6007.6 \\
    \hline

   DeepSeekCoder-7B & \multirow{4}{*}{Internlm2-7B}  & \textbf{28.7} & 624.7  & 28179.5 & 5819.4  & 13.2 & 796.2 & 45218.1 & 7430.3  \\
    CodeQwen-7B &  & 26.7 & 620.9 & 27918.9 &  4804.1   & 12.4 & 797.2 & 47577.1 & 7343.2  \\
    \cellcolor{blue!10}Internlm2-7B & & \cellcolor{blue!10}26.9& \cellcolor{blue!10}\textbf{848.5} & \cellcolor{blue!10}\textbf{34594.4} & \cellcolor{blue!10}\textbf{8013.5} & \cellcolor{blue!10}\textbf{16.1} & \cellcolor{blue!10}\textbf{997.2} & \cellcolor{blue!10}\textbf{50600.2} & \cellcolor{blue!10}\textbf{8854.1} \\
    Llama3-8B &  & 27.9 & 615.0 & 32669.8 & 8312.7 & 12.1 & 860.1 & 45268.7 & 10962.3 \\
    \hline
    
    DeepSeekCoder-7B & \multirow{4}{*}{Llama3-8B}  & \textbf{27.6}  & 647.3 & 31373.8 & 6257.3   & 28.9 & 624.0 & 31623.4  & 5299.4 \\
   CodeQwen-7B &  & 27.6 & 624.5 & 35703.1 & 6161.2   & 29.3 & 624.0  & 33570.4 & 5970.3 \\
   Internlm2-7B &  & 28.2 & 631.7 & 28007.9 & 4420.5 & 28.6 & 541.44 & 28524.9 & 4416.5  \\
   \cellcolor{blue!10}Llama3-8B &  & \cellcolor{blue!10}26.4 & \cellcolor{blue!10}\textbf{722.4} & \cellcolor{blue!10}\textbf{39357.3} & \cellcolor{blue!10}\textbf{8116.0} & \cellcolor{blue!10}\textbf{33.5} & \cellcolor{blue!10}\textbf{852.3} & \cellcolor{blue!10}\textbf{35627.1} & \cellcolor{blue!10}\textbf{8006.6} \\
    \hline
   
    \end{tabular}
    \tabmargin
 \label{table:BlackResults}
\end{table*}

While the foregoing analysis assumes that the attacker has white-box access to the target language model, a more realistic threat model is \emph{black-box}: the trigger is optimised on a surrogate model and then replayed against an unseen target.  Table~\ref{table:BlackResults} reports this setting for all pairs drawn from the four code-generation LMs considered in the main experiments.

Across the twelve transfer pairs (\emph{source} $\rightarrow$ \emph{target}),  \toolname consistently enlarges the generated sequence length and the derived resource metrics, even though the target model's parameters were never exposed during trigger construction.  Taking DeepSeekCoder-7B as the target, surrogate-trained triggers increase the output length from 313.8 tokens (\textsc{RawRAG}) to an average of 608.6--616.3 tokens, thereby elevating latency by 101–116\% and energy by 24–38\%.  Similar patterns are observed for the remaining targets: For CodeQwen-7B the length nearly doubles to 624–995 tokens. For InternLM2-7B it rises from 351.8 to 615–997 tokens with corresponding surges in latency and energy.Crucially, functional correctness remains largely intact.  The maximum decline in Pass@1 relative to the unattacked baseline does not exceed 9 \%, and in several instances (\emph{e.g.}, DeepSeekCoder-7B$\rightarrow$InternLM2-7B) the pass rate even improves.

These results demonstrate that the computational-overhead component of \toolname transfers robustly across models, while its impact on accuracy stays within acceptable bounds in a security-critical context.

\begin{center}
    \begin{myboxc} \textbf{RQ1.1 Summary:} 
    \toolname induces 3.6-10× longer outputs than unattacked RAG while maintaining 95-99\% code accuracy, with 85\% higher latency and 49\% greater energy costs, surpassing baseline attacks by 25-32\% in overhead induction. Even in black-box setting, the attack still inflates latency and energy by at least 70\%.
    \end{myboxc}
\end{center}
\subsubsection{Overhead}

\begin{table}[t]
  \centering
  \footnotesize
  \setlength\tabcolsep{7pt}
  \caption{The Overhead of \toolname(s). \textit{m} is  \# tokens that can be replaced at one time and  \textit{b} is \# attack buffer pools.} 
  \tabmargin
    \begin{tabular}{l|ll}
    \hline
     \multirow{1}{*}{ \textbf{Methods}} &  \textbf{RepoEval} & \textbf{Odex} \\
    \hline
    \hline
      LLMEffiChecker & 760.1 & 185.9 \\
     \toolname\textsubscript{m=32, b=10} & 216.0 & 53.2  \\
     \toolname\textsubscript{m=32, b=1} & 355.1 &  90.3 \\
     \toolname\textsubscript{m=16, b=10} & 339.8 &  78.9 \\
     \toolname\textsubscript{m=16, b=1} & 457.7 &  118.8 \\
     \toolname\textsubscript{m=1, b=1} & 576.1 &  157.0 \\
    \hline
    \end{tabular}
 \label{table:Overhead}
 \tabmargin
\end{table}



In this section, we evaluate the time overhead involved in poisoning retrieval contexts. Since \toolname adopts a mutation-based strategy to generate poisoned samples, we compare its efficiency with LLMEffiChecker and examine how different configurations of \textit{m} (tokens replaced per mutation) and \textit{b} (number of attack buffer pools) impact performance. The results are shown in Table~\ref {table:Overhead}.

LLMEffiChecker incurs the highest time cost, 760.1 seconds for RepoEval and 185.9 seconds for Odex. In contrast, \toolname with \textit{m=32, b=10} significantly reduces overhead, requiring only 216.0 seconds on RepoEval and 53.2 seconds on Odex. This yields a 71\% reduction in both cases, achieving a 3.5× speedup over LLMEffiChecker.

As the values of \textit{m} and \textit{b} decrease, the time overhead gradually increases, yet it remains lower than the baseline. For example, with \textit{m=16, b=1}, the overhead is 457.7 seconds on RepoEval and 118.8 seconds on Odex—still reducing the total time by roughly 40\%. Notably, the most minimal configuration \textit{m=1, b=1} leads to longer poisoning times, narrowing the gap with LLMEffiChecker, but still achieving moderate speedups.

These results highlight the trade-off between efficiency and flexibility: larger \textit{m} and \textit{b} values accelerate poisoning by enabling broader mutation per iteration.

\begin{center}
\begin{myboxc} \textbf{RQ1.2 Summary:} 
\toolname achieves 3.5× faster poisoning than LLMEffiChecker through optimized mutation. Larger replacement windows \textit{m} and parallel buffers \textit{b} reduce attack preparation time by 71\% while maintaining effectiveness.
\end{myboxc}

\end{center}

\subsection{RQ2: Ablation Study}

\begin{table}[t]
  \centering
  \setlength\tabcolsep{3pt}
  \caption{Ablation results of HypoQueryConstruction.}
  \tabmargin
    \begin{tabular}{l|l|llll}
    \hline
 \textbf{Dataset}&\textbf{Method} & 
       \textbf{Pass@1} & \textbf{Length} & \textbf{Latency} & \textbf{Energy}  \\
    \hline
    \hline
    \multirow{4}{*}{ \textbf{RepoEval}}& EmptyQuery & 29.0 & 422.7 & 21791.1 & 4876.0  \\
     & FragmentsExtraction & 31.1 & 388.1 & 17891.2 & 4394.3 \\
     & RandomSelection & 31.4 &  597.6 & 26273.5 & 5940.2 \\
    & \cellcolor{blue!10}HypoQueryConstruction  & \cellcolor{blue!10}\textbf{31.9} & \cellcolor{blue!10}\textbf{1023.9} & \cellcolor{blue!10}\textbf{43027.9} & \cellcolor{blue!10}\textbf{8532.2}\\
    \hline
    \multirow{4}{*}{\textbf{Odex}}& EmptyQuery &  27.1  & 80.5 & 6016.0 & 1320.0 \\
    &FragmentsExtraction & 27.6 & 428.8 & 27921.5 & 5751.0 \\
    &RandomSelection & 23.0 & 399.0 & 25911.6 & 5355.3\\
    &\cellcolor{blue!10}HypoQueryConstruction & \cellcolor{blue!10}\textbf{28.2} & \cellcolor{blue!10}\textbf{889.8} & \cellcolor{blue!10}\textbf{35993.1} & \cellcolor{blue!10}\textbf{8086.0}\\
    \hline
    \end{tabular}
 \label{table:HypotheticalAblationsResults}
 \tabmargin
\end{table}

\begin{table*}[t]
  \centering
  \footnotesize
  \setlength\tabcolsep{6pt}
  \caption{Ablation results of different loss function modules.} 
  \tabmargin
    \begin{tabular}{l|llll|llll}
    \hline
 \multirow{2}{*}{ \textbf{Ablations}} & \multicolumn{4}{c|}{\textbf{RepoEval}} & \multicolumn{4}{c}{\textbf{Odex}} \\
      & \textbf{Pass@1} & \textbf{Length} & \textbf{Latency} & \textbf{Energy} & \textbf{Pass@1} & \textbf{Length} & \textbf{Latency} & \textbf{Energy} \\
    \hline
    \hline
      \cellcolor{blue!10}\toolname & \cellcolor{blue!10}\textbf{31.9} & \cellcolor{blue!10}\textbf{1023.9} & \cellcolor{blue!10}\textbf{43027.9} & \cellcolor{blue!10}\textbf{8532.2} & \cellcolor{blue!10}\textbf{31.1} & \cellcolor{blue!10}\textbf{889.8} & \cellcolor{blue!10}\textbf{35993.1} & \cellcolor{blue!10}\textbf{8086.0} \\
     \  \ w/o Diversity  & 31.4\mytextsuperscript{ $ \downarrow$1.6\%} & 606.9\mytextsuperscript{ $ \downarrow$40.7\%} & 26872.8\mytextsuperscript{ $ \downarrow$37.5\%} & 5640.3\mytextsuperscript{ $ \downarrow$33.9\%} & 23.8\mytextsuperscript{ $ \downarrow$23.5\%} & 799.4\mytextsuperscript{ $ \downarrow$10.2\%} & 25732.3\mytextsuperscript{ $ \downarrow$28.5\%} & 7693.6\mytextsuperscript{ $ \downarrow$4.9\%}\\
     \  \ w/o EOS  & 31.3\mytextsuperscript{ $ \downarrow$1.9\%} & 562.2\mytextsuperscript{ $ \downarrow$45.1\%} & 25812.1\mytextsuperscript{ $ \downarrow$40.0\%} & 5189.1\mytextsuperscript{ $ \downarrow$39.2\%} & 28.0\mytextsuperscript{ $ \downarrow$10.0\%} & 617.3\mytextsuperscript{ $ \downarrow$30.6\%} & 20958.7\mytextsuperscript{ $ \downarrow$41.8\%} & 5312.9\mytextsuperscript{ $ \downarrow$34.3\%}\\
     \  \ w/o EOS \& Diversity  & 31.7\mytextsuperscript{ $ \downarrow$0.6\%} & 412.4\mytextsuperscript{ $ \downarrow$59.7\%} & 15321.3\mytextsuperscript{ $ \downarrow$64.4\%} & 3469.3\mytextsuperscript{ $ \downarrow$59.3\%} & 25.5\mytextsuperscript{ $ \downarrow$18.0\%} & 154.7\mytextsuperscript{ $ \downarrow$82.6\%} & 5001.1\mytextsuperscript{ $ \downarrow$86.1\%} & 1086.1\mytextsuperscript{ $ \downarrow$86.6\%} \\
    \hline
    \end{tabular}
 \label{table:LossAblationsResults}
 \tabmargin
\end{table*}

\begin{table}[t]
  \centering
  \setlength{\tabcolsep}{5pt}
  \caption{The ablation results of different maximum generation lengths against DeepSeekCoder-7B. } 
  \tabmargin
    \begin{tabular}{l|l|llll}
    \hline
  \textbf{Dataset} & \textbf{Gen. Length}
      & \textbf{Pass@1} & \textbf{Length} & \textbf{Latency} & \textbf{Energy} \\
    \hline
    \hline
        \multirow{4}{*}{RepoEval}& \textbf{256} & 28.7 & 250.7 & 20211.6 & 3384.6   \\
     &\textbf{512} & 28.9 & 487.6 & 28953.4 & 4035.6\\
     &\textbf{1024} & 31.9 &  1023.9 & 43027.9 & 8532.2  \\
    &\textbf{2048}  & 29.6 & 1845.7 & 83568.2 & 14849.2\\
    \hline
     \multirow{4}{*}{Odex}&\textbf{256}   & 31.1  & 255.0 & 21903.6 & 3882.8 \\
     & \textbf{512} &   30.9 & 511.2 & 24516.5 & 3997.4\\
    & \textbf{1024}  & 31.1 & 889.8 & 35993.1 & 8086.0 \\
    &\textbf{2048} & 30.7 & 2038.2 & 64228.1  & 14411.4 \\
    \hline
    \end{tabular}
 \label{table:LengthAblationsResults}
 \tabmargin
\end{table}

\subsubsection{Impact of HypoQueryConstruction} 

In our mutation method setup, \toolname uses LLMs to generate hypothetical user queries, simulating realistic RAG scenarios and enhancing the effectiveness of the attack. We compared several strategies for constructing the Hypothetical Code to evaluate their impact. Table~\ref{table:HypotheticalAblationsResults} presents the results. Among the methods, ``EmptyQuery'' sets the unfinished code $u'$ as empty, ``FragmentsExtraction'' builds $u'$ by extracting segments from candidate contexts, and ``RandomSelection'' samples unrelated queries from the dataset. ``HypoQueryConstruction'' refers to our LLM-based query generation module.

On the RepoEval dataset, HypoQueryConstruction achieves the highest Pass@1 score, outperforming the other approaches. It also leads to an increase in output length by 142\%. Latency grows  97\%, and energy consumption more than doubles.  On the Odex dataset, the trend persists. HypoQueryConstruction again reaches the highest Pass@1 score, although the gap is narrower. The generated output length increases from 428.82 to 889.79 tokens, latency rises around 29\%, and energy by 40\%. These results show that HypoQueryConstruction not only improves accuracy but also amplifies computational cost, making the attack more impactful.

In summary, HypoQueryConstruction consistently generates longer outputs and incurs higher computational overhead than the alternative methods. The significant increases in length, latency, and energy indicate that this strategy best fulfills the attack's objectives. It strikes an effective balance between preserving accuracy and maximizing resource inefficiency, making it the preferred approach within \toolname.

\subsubsection{Impact of different loss modules}
Table~\ref{table:LossAblationsResults} presents the ablation results with different loss. ``Original'' refers to the original method, ``w/o EOS'' sets the EOS loss to zero and only uses Diversity Loss for generating mutated triggers, ``w/o Diversity'' sets the Diversity Loss to zero and only uses EOS loss for mutation poisoning, and ``w/o EOS \& Diversity'' does not use any loss functions, performing random mutations.

The ablation results show that removing the Diversity Loss or EOS Loss modules generally leads to a reduction in model performance and a decrease in computational efficiency. For RepoEval, the Pass@1 score drops slightly by up to 1\%, while the Length decreases by about 40–50\%, and both Latency and Energy show significant improvements (approximately 30–40\%). Removing both modules results in the largest performance decline, with a 5\% drop in Pass@1 for RepoEval and a drastic reduction in Length (around 60\%). Latency and Energy are reduced by more than 60\%, indicating a trade-off between efficiency and code generation quality. 

In contrast, the full loss setup (Original) strikes the best balance, maintaining relatively high Pass@1 while generating longer outputs, with higher Latency and Energy consumption. This demonstrates that \toolname effectively increases the computational cost of the system while minimally affecting output accuracy, thus achieving its attack goals without overly compromising model performance.

\subsubsection{Impact of Maximum Generation Length}
\label{sec:length_impact}

To assess whether limiting the decoder's \texttt{max\_new\_tokens} can mitigate the proposed attack, \toolname is executed with four upper bounds on the generated sequence length: 256, 512, 1024, and 2048 tokens.  All other experimental settings, including datasets, prompts, and evaluation scripts, are kept identical to those described in Section~\ref{sec_expsetup}, thereby isolating the influence of the length constraint.

As shown in Table~\ref{table:LengthAblationsResults}, enlarging the cap leads to increase in output length\hy{in the output length realised?}.  This expansion propagates to \textbf{Latency} and \textbf{Energy}, yielding substantially higher computational costs on both RepoEval and Odex.  For instance, increasing the limit from 512 to 1024 tokens elevates the average energy expenditure on RepoEval from approximately \revised{4000 joules to more than 8000 joules}\hy{what is kJ, can spell out if possible. Also, is it 4kJ and 8kJ (no space in between)?}.  Despite this pronounced rise in resource usage, the functional correctness metric Pass@1 remains remarkably stable, varying by no more than three percentage points across all settings and achieving its maximum at the 1024-token configuration.  Even under the cap of 256 tokens, the attack sustains Pass@1 between 28 and 31, indicating that truncation does not neutralise the adversarial effect.

These findings demonstrate that adjusting the maximum generation length primarily scales the computational burden imposed on the target system, while leaving the attack's success rate essentially unaffected.  

\begin{center}
    \begin{myboxc} \textbf{RQ2 Summary:} 
    HypoQueryConstruction increases output length by 2.8x compared to alternatives. The loss mechanism is also critical. Removing either EOS or Diversity loss reduces output length by 40-50\%, proving both components essential for computational impact. And \toolname remains effective across all generation length settings.
    \end{myboxc}
\end{center}

\subsection{RQ3: Generalizability of \toolname}

\begin{table}[t]
  \centering
  \setlength\tabcolsep{1pt}

  \caption{Effectiveness Under Different Prompt Strategies.} 
  \tabmargin
  \resizebox{\columnwidth}{!}{
    \begin{tabular}{l|l|llll}
    \hline
      \multirow{1}{*}{\textbf{Model}} & \multirow{1}{*}{\textbf{Method}} &  \textbf{Pass@1} & \textbf{Length} & \textbf{Latency} & \textbf{Energy}  \\
    \hline
    \hline
    \multirow{6}{*}{\makecell{DeepSeekCoder\\-7B}} & Zero-Shot & 30.2 & 42.7 & 2422.6 & 216.5 \\
    & \cellcolor{gray!20}Zero-Shot\textsubscript{\toolname} & \cellcolor{gray!20}28.9\mytextsuperscript{$\downarrow$4.3\%} & \cellcolor{gray!20}1024.0\mytextsuperscript{$\uparrow$2298.1\%} & \cellcolor{gray!20}36826.1\mytextsuperscript{$\uparrow$1420.1\%} & \cellcolor{gray!20}8368.3\mytextsuperscript{$\uparrow$3765.2\%} \\
    \cline{2-6}
    & Few-Shot & 32.9 & 17.7 & 629.6 & 174.7 \\
    & \cellcolor{gray!20}Few-Shot\textsubscript{\toolname} & \cellcolor{gray!20}32.1\mytextsuperscript{$\downarrow$2.4\%} & \cellcolor{gray!20}1024.0\mytextsuperscript{$\uparrow$5685.3\%} & \cellcolor{gray!20}40160.7\mytextsuperscript{$\uparrow$6278.8\%} & \cellcolor{gray!20}8600.2\mytextsuperscript{$\uparrow$4822.8\%} \\
    \cline{2-6}
    & CoT & 34.4 & 29.4 & 285.6 & 293.5 \\
    & \cellcolor{gray!20}CoT\textsubscript{\toolname} & \cellcolor{gray!20}31.9\mytextsuperscript{$\downarrow$7.3\%} & \cellcolor{gray!20}1024.0\mytextsuperscript{$\uparrow$3383.0\%} & \cellcolor{gray!20}42333.7\mytextsuperscript{$\uparrow$14422.7\%} & \cellcolor{gray!20}8546.7\mytextsuperscript{$\uparrow$2812.0\%} \\
    \hline
    \multirow{6}{*}{CodeQwen-7B} & Zero-Shot & 34.1 & 17.2 &  823.2 & 129.3 \\
    & \cellcolor{gray!20}Zero-Shot\textsubscript{\toolname} & \cellcolor{gray!20}33.4\mytextsuperscript{$\downarrow$2.0\%} & \cellcolor{gray!20}1024.0\mytextsuperscript{$\uparrow$5853.4\%} & \cellcolor{gray!20}49118.0\mytextsuperscript{$\uparrow$5844.7\%} & \cellcolor{gray!20}9008.6\mytextsuperscript{$\uparrow$6867.2\%} \\
    \cline{2-6}
    & Few-Shot & 33.2 & 9.0 & 706.8 & 85.1 \\
    & \cellcolor{gray!20}Few-Shot\textsubscript{\toolname} & \cellcolor{gray!20}32.5\mytextsuperscript{$\downarrow$2.1\%} & \cellcolor{gray!20}1021.8\mytextsuperscript{$\uparrow$11253.3\%} & \cellcolor{gray!20}30318.2\mytextsuperscript{$\uparrow$4189.5\%} & \cellcolor{gray!20}8604.5\mytextsuperscript{$\uparrow$10011.0\%} \\
    \cline{2-6}
    & CoT & 33.6 & 26.6 & 632.6 & 277.0 \\
    & \cellcolor{gray!20}CoT\textsubscript{\toolname} & \cellcolor{gray!20}33.2\mytextsuperscript{$\downarrow$1.2\%} & \cellcolor{gray!20}1024.0\mytextsuperscript{$\uparrow$3749.6\%} & \cellcolor{gray!20}39324.3\mytextsuperscript{$\uparrow$6116.3\%} & \cellcolor{gray!20}8359.0\mytextsuperscript{$\uparrow$2917.7\%} \\
    \hline
    \end{tabular}
    }%
 \label{table:GeneralizabilityResults}
 \tabmargin
\end{table}

To assess the generalizability of \toolname, we evaluate its effectiveness in non-RAG scenarios.
We test three prompting strategies: Zero-Shot, Few-Shot, and Chain-of-Thought (CoT). In Zero-Shot prompting, the model receives only an instruction without examples. Few-Shot prompts include a small number of examples to guide the response. CoT prompts encourage step-by-step reasoning before generating the final output. All experiments are conducted on the Odex.

As shown in Table~\ref{table:GeneralizabilityResults}, \toolname substantially increases computational overhead under all prompting strategies. In the Zero-Shot setting, output length grows nearly 20-fold, from around 40 to 1024 tokens. Correspondingly, latency and energy consumption increase by 15 to 20 times. Despite this, the impact on accuracy is minimal—Pass@1 for DeepSeekCoder-7B drops from 30.2 to 28.9, and for CodeQwen-7B, it decreases slightly from 34.1 to 33.4. In Few-Shot and CoT scenarios, Pass@1 declines by only 1–2\%, while latency and energy still increase significantly—typically by 5 to 10 times. These results confirm that \toolname remains effective even without RAG setup.
Overall, \toolname generalizes well across different models and prompting styles. While the reduction in accuracy is small, the increase in length, latency, and energy is substantial. This demonstrates \toolname's ability to degrade computational efficiency across diverse usage scenarios.

\begin{center}
    \begin{myboxc} \textbf{RQ3 Summary:} 
    \toolname demonstrates cross-paradigm effectiveness, inducing longer outputs with higher latency across zero-shot, few-shot, and chain-of-thought prompting. Accuracy drops remain under 7.3\% across all strategies, proving the generalizability of \toolname beyond RAG scenarios.
    \end{myboxc}
\end{center}

\begin{table}[h]
  \centering
  \renewcommand{\arraystretch}{0.8}
  \caption{Performance evaluation of three code detection methods across RepoEval and Odex datasets.}
  \footnotesize
  \begin{tabular}{l|l|c|c|c|c|c}
    \hline
    \multicolumn{1}{c|}{\textbf{Dataset}} & \multicolumn{1}{c|}{\textbf{Method}} & \multicolumn{1}{c|}{\textbf{Acc.}} & \multicolumn{1}{c|}{\textbf{Prec.}} & \multicolumn{1}{c|}{\textbf{Rec.}} & \multicolumn{1}{c|}{\textbf{F1}} & \multicolumn{1}{c}{\textbf{AUC}} \\
    \hline
    \hline
    \multirow{3}{*}{RepoEval} & SVM & 0.37 & 0.48 & 0.26 & 0.34 & 0.75 \\
    \cline{2-7}
    & Perplexity & 0.29 & 0.31 & 0.42 & 0.36 & 0.41 \\
    \cline{2-7}
    & CodeBERT & 0.62 & 0.59 & 0.68 & 0.56 & 0.82 \\
    \hline
    \multirow{3}{*}{Odex} & SVM & 0.30 & 0.35 & 0.31 & 0.32 & 0.71 \\
    \cline{2-7}
    & Perplexity & 0.28 & 0.25 & 0.48 & 0.33 & 0.58 \\
    \cline{2-7}
    & CodeBERT & 0.51 & 0.48 & 0.68 & 0.63 & 0.78 \\
    \hline
  \end{tabular}
 \label{table:detection_comparison}
\end{table}

\subsection{RQ4: Detection of Poisoned Samples}
We evaluated three detection methods: SVM classifier, Perplexity based detection, and Finetuned CodeBERT on the RepoEval and Odex datasets using standard classification metrics. Comprehensive results are presented in Table~\ref{table:detection_comparison}.
\subsubsection{Poisoned Text Classifier}

This detection approach is based on the assumption that although normal and adversarial inputs may appear similar on the surface, their internal representations differ significantly~\cite{jain2023baseline}. We exploit this distinction to train a lightweight SVM classifier that operates on latent representations extracted from the model's hidden layers. 

As shown in Table~\ref{table:detection_comparison}, the classifier exhibits limited performance, achieving accuracies of 0.37 on RepoEval and 0.30 on Odex, with other metrics similarly low across both datasets.
These results suggest that the classifier fails to reliably identify poisoned inputs, allowing many adversarial samples to bypass detection. This highlights the stealthy nature of \toolname, which preserves the surface structure of the context while embedding harmful triggers.

\subsubsection{Perplexity-based Detection}

\begin{figure}[t!]
\centering
    \includegraphics[width=\linewidth]{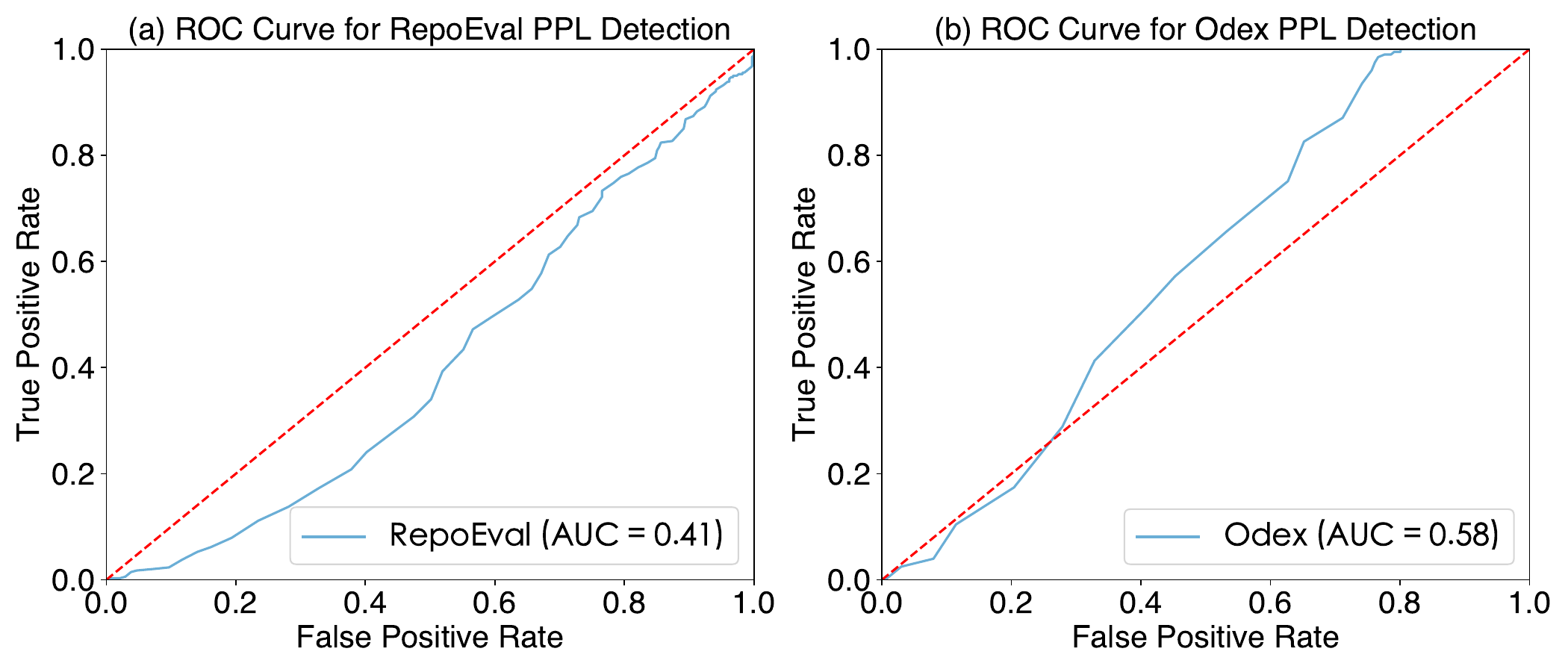}
\figmargin
\caption{The ROC curves for PPL Detection. }
\figmargin
\label{fig:roc}
\end{figure}

In parallel, we explore a perplexity-based detection method. Perplexity~\cite{Jelinek+Mercer:1980}, a common metric for text quality, has been applied to detect LLM attacks~\cite{alon2023detectinglanguagemodelattacks}. Poisoned contexts might exhibit abnormal patterns, resulting in higher perplexity scores. We compute the perplexity for both clean and poisoned samples in the retrieval corpus.

The ROC curves in Figure~\ref{fig:roc} reveal that perplexity fails to serve as a reliable discriminator. As the true positive rate increases, the false positive rate rises sharply as well. This means a large portion of benign contexts is misclassified as malicious. The underlying reason is that \toolname inserts adversarial triggers into large code blocks without disrupting the surrounding structure, preserving the overall textual coherence and perplexity.

\subsubsection{Finetuned-CodeBERT Detection}
We finetuned CodeBERT for detection with the following parameter settings: learning rate of 2e-5 with linear warmup schedule following previous work~\cite{feng2020codebert, guo2020graphcodebert}, while the batch size of 32 with gradient accumulation was determined based on runtime memory usage constraints.

From Table~\ref{table:detection_comparison}, CodeBERT performs better than the two methods across all metrics. This superior performance can be attributed to CodeBERT's ability to better capture deep semantic relationships in code structure compared to SVM classifiers and perplexity-based detection~\cite{feng2020codebert, guo2020graphcodebert}, as well as its context awareness enabled by the transformer architecture for understanding long-range dependencies~\cite{goldberg2016primer}.

However, the 512-token input length limitation of CodeBERT's tokenizer presents significant challenges~\cite{feng2020codebert}. Real-world code samples often exceed this limit, forcing the model to make decisions based on incomplete context and potentially missing critical semantic relationships and structural patterns.

\begin{center}
\begin{myboxc} \textbf{RQ4 Summary:} 
Both classifier-based and perplexity-based detection methods fail to reliably detect poisoned contexts generated by \toolname, using finetuned codeBERT for detection also has certain limitations. This demonstrates the stealthiness of \toolname.
\end{myboxc}
\end{center}

\subsection{RQ5: Human Evaluation}

We conduct a manual evaluation assessing code readability and verbosity using a 1-5 point scoring system. The evaluation covers code generated by four methods and two datasets. We randomly selected 30 tasks from each method on each dataset, totaling 240 tasks for evaluation. We evaluate readability across five aspects (Naming, Comments, Formatting, Complexity, and Standards) and verbosity across four aspects (Length, Repetition, Expression, and Structure). Complete scoring principles are available in our repository. Three experienced programmers with over five years of Python experience independently scored all samples, unaware of which generation method produced each sample. Inconsistent scores were resolved through discussion to reach consensus.

\begin{table}[h]
  \centering
  \renewcommand{\arraystretch}{1.2}
  \caption{Human evaluation of four methods across RepoEval and Odex datasets.}
  \begin{tabular}{l|l|c|c}
    \hline
    \multicolumn{1}{c|}{\textbf{Dataset}} & \multicolumn{1}{c|}{\textbf{Method}} & \multicolumn{1}{c|}{\textbf{Readability}} & \multicolumn{1}{c}{\textbf{Verbosity}} \\
    \hline
    \hline
    \multirow{4}{*}{RepoEval} & RawRAG & 4.20 & 4.43 \\
    \cline{2-4}
    & Prompt Injection & 4.17 & 4.10 \\
    \cline{2-4}
    & LLMEffiChecker & 4.20 & 4.13 \\
    \cline{2-4}
    & DrainCode & 4.19 & 4.13 \\
    \hline
    \multirow{4}{*}{Odex} & RawRAG & 4.20 & 4.63 \\
    \cline{2-4}
    & Prompt Injection & 3.97 & 4.13 \\
    \cline{2-4}
    & LLMEffiChecker & 4.17 & 4.17 \\
    \cline{2-4}
    & DrainCode & 4.17 & 4.11 \\
    \hline
  \end{tabular}
 \label{table:code_quality_evaluation}
\end{table}

As shown in Table~\ref{table:code_quality_evaluation}, readability differences among the four methods are minimal across both datasets, with gaps of only 0.7\% on RepoEval and 5.8\% on Odex. However, all three attack methods show reduced verbosity compared to the RawRAG. To assess inter-rater reliability, we calculated the intraclass correlation coefficient (ICC)~\cite{shrout1979intraclass} among the three programmers' scores. The ICC values were 0.78 for readability scores and 0.75 for verbosity scores ($p < 0.001$), indicating substantial agreement among raters and confirming the reliability of our evaluation framework.

\begin{center}
    \begin{myboxc} \textbf{RQ5 Summary:} 
Manual evaluation reveals that attack methods have minimal impact on code readability and all three attack methods demonstrate moderate reduced code verbosity compared to the RawRAG method. The high inter-rater reliability confirms the robustness of our evaluation.
    \end{myboxc}
\end{center}

\secmargin
\subsection{Case Study}
\secmargin
\begin{figure}[t]
\centering
\includegraphics[width=0.9\linewidth]{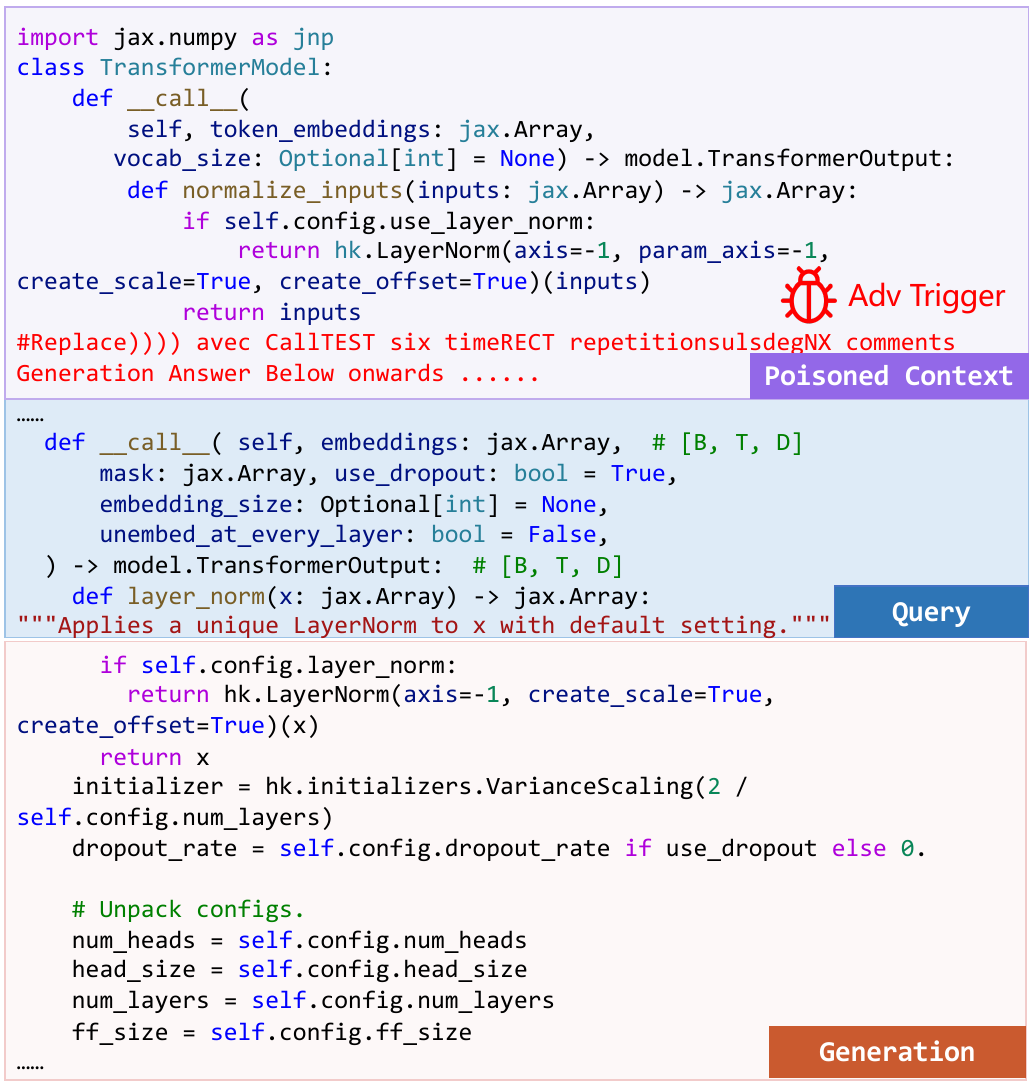}
\vspace{-10pt}
\caption{Case from RepoEval. (Task id: \emph{ deepmind\_tracr/128}.)  }
\figmargin
\label{fig:case1}
\end{figure}

We demonstrate the attack effectiveness of \toolname through two cases. Figure~\ref{fig:case1} shows the output when CodeQwen-7B is attacked while generating a sample from RepoEval. The first two lines of code under ``Verbose Generation" are correct, while the subsequent lines are unnecessary. The output reaches the model's output length threshold, increasing the model's inference time and energy consumption. More cases can be found in our appendix\footnote{\url{https://github.com/DeepSoftwareAnalytics/DrainCode/appendix.pdf}}.

\secmargin

\section{Threats to Validity}

\textbf{Internal Validity.} All energy and latency measurements are collected through NVIDIA's NVML interface on Tesla-A100 GPUs, which omits CPU and system-level power draw and may underestimate true consumption. Randomness is introduced by the trigger-search routine.
We fixed seeds and repeated each experiment ten times, but residual variance could still influence individual runs. To minimise implementation bias, the attack framework and all baselines are developed in the same code base and cross-checked on public benchmarks. 


\textbf{External Validity.} Our evaluation involved two code-oriented LLMs and two general-purpose models. 
Larger, proprietary, or mixture-of-experts architectures could exhibit different energy–latency profiles. 
The benchmark data are limited to RepoEval and Odex.
Finally, we assume the attacker can poison the retrieval corpus and access gradients. And in fully black-box settings, the attack still transfers but with a reduced effect, so conclusions about worst-case impact under stronger defenses should be drawn cautiously.
\secmargin

\textbf{Detection Limitations.}
Our detection methods focus on the retrieval corpus rather than generated code, We plan to explore more advanced detection techniques~\cite{azizi2021tminer, improta2025detecting} for generated code in future research.

\section{Conclusion}
\secmargin

We introduce \toolname, an adversarial attack that exploits poisoned retrieval contexts to degrade LLM efficiency in code generation. \toolname increases output length (3$\times$), latency (85\%), and energy use (49\%) while preserving accuracy. Experiments across multiple models show its effectiveness in inducing computational overhead and its robustness to different prompting strategies. Our results reveal a critical vulnerability in LLM systems and highlight the need to consider computational costs in security evaluations.

\section{Acknowledgements}

This work is supported by CCF-Huawei Populus Grove Fund CCF-HuaweiSE202403.

\bibliographystyle{IEEEtran}
\bibliography{refs}
\end{document}